# NONLINEAR PREDICTION WITH NEURAL NETS IN ADPCM

*Marcos Faundez-Zanuy* , *Francesc Vallverdú***, *Enric Monte***
*Escola Universitària Politècnica de Mataró
**Signal Theory & Communications Department (UPC)
Avda. Puig i Cadafalch 101-111, E-08303 Mataró (BARCELONA)
e-mail: faundez@eupmt.upc.es http://www.eupmt.upc.es/veu

## ABSTRACT
In the last years there has been a growing interest for nonlinear speech models. Several works have been published revealing the better performance of nonlinear techniques, but little attention has been dedicated to the implementation of the nonlinear model into real applications. This work is focused on the study of the behaviour of a nonlinear predictive model based on neural nets, in a speech waveform coder. Our novel scheme obtains an improvement in SEGSNR between 1 and 2 dB for an adaptive quantization ranging from 2 to 5 bits.

## 1. INTRODUCTION
In the last decade several studies dealing with the nonlinear prediction of speech have been reported. Most part of the bibliography has been focused on parametric prediction based on neural nets, because they are the approach that offers the best improvement over LPC analysis.

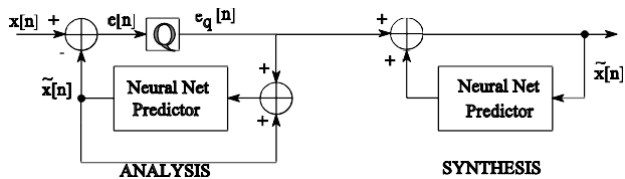

Fig. 1 ADPCM closed loop structure

In this paper we propose a novel ADPCM (see fig.1) speech waveform coder for the following bit rates: 16Kbps, 24Kbps, 32Kbps and 40Kbps with
a) nonlinear predictor
b) hybrid (linear/nonlinear) predictor (see fig. 9)

## 2. ADPCM WITH NONLINEAR PREDICTOR SCHEME
In order to compare the nonlinear speech prediction system, ADPCM waveform coder is used. The nonlinear predictor is compared against the traditional LPC one, with the following characteristics:

### 2.1 System overview
*Predictor coefficients updating*

! The coefficients are updated once time every frame.

! To avoid the transmission of the predictor coefficients an ADPCM backward (ADPCMB) configuration is adopted. That is, the coefficients of the predictor are computed over the decoded previous frame, because it is already available at the receiver and it can compute the same coefficients values without any additional information. The obtained results with a forward unquantized predictor coefficients (ADPCMF) are also provided for comparison purposes.

! The nonlinear analysis consists on a multilayer perceptron with 10 input neurons, 2 hidden neurons and 1 output neuron. The network is trained with the Levenberg-Marquardt algorithm.

! The linear prediction analysis of each frame consists on an all-pole filter, 10 coefficients obtained with the autocorrelation method (LPC-10) and 25 order filter (LPC-25).
*Residual prediction error quantization*

! The prediction error has been quantized with 2 to 5 bits. (bit rate from 16Kbps to 40Kbps).

! The quantizer step is adapted with multiplier factors, obtained from [1]. $\Delta_{max}$ and $\Delta_{min}$ are set empirically.
*Database*

! The results have been obtained with the following database: 8 speakers (4 males & 4 females) sampled at 8Khz and quantized at 12 bits/sample.
Additional details about the predictor and the database were reported in [2].

### 2.2 Parameter selection
a)Linear predictor
For the linear predictor the parameters are:

! Prediction order: it is studied LPC-10 (same number of input samples than the MLP 10x2x1) and LPC-25 (same number of prediction coefficients than the MLP 10x2x1

! Frame length: sizes from 10 to 300 samples with a step of 10 samples are evaluated. Obviously, the bigger frame size implies a smaller number of frames for a given speech signal, so the computational complexity is reduced, but if the frame length is very large then the assumption of stationary signal into the analysis window is no valid and the behaviour is degraded. If the frame length is small, the parameter estimation is not robust enough and the behaviour degrades.

b)Nonlinear predictor
For the nonlinear predictor based on neural nets, the number of parameters that must be optimized is greater. The selected network architecture is the Multi-Layer Perceptron with 10 input neurons, 2 hidden neurons with a sigmoid transfer function and one output neuron with a linear transfer function trained with the Levenberg-Marquardt algorithm, based on our previous results [2]. The adjusted parameters of the predictor into the closed loop ADPCM scheme are:

! Number of trained epochs: This is a critical parameter. To encode a given frame the neural net is trained over the previous frame in the backward scheme and over the actual frame in the forward configuration. In both cases special attention must be taken in order to avoid the problem of overtraining (the network must have a good generalization capability to manage inputs not used for training). Although consecutive frames are normally very similar, there are significative changes in the waveform that must be seen as perturbances of the input, and even if the neural net is applied over the same frame used for training, the conditions are different because the predictor is trained in an open-loop scheme and tested in closed loop, so really the input signal is corrupted by the quantization noise. This is as much important the lesser is the number of quantizer bits. The way to make the neural net as robust as possible to this small changes implies the optimization of training conditions such us:

a)Number of epochs used for training

b)Number of random initializations of the weights ( a multistart algorithm is used).

For showing the importance of this subject, the following experiment was done: over a voiced portion of one speech signal two consecutive frames were selected. Figure 2 represents on the top the frame used for training (for the forward scheme) and on the bottom the next frame (used for testing, and for the backward configuration).

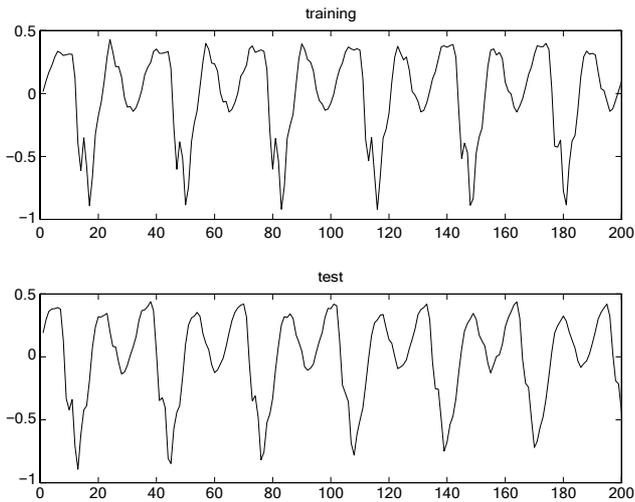

Figure 2. Frames used for training and testing.

It is known that nets with a low number of weights are proved to fall in local minimums. In order to study empirically the properties of the multistart methods, we show the results of three experiments in figures 3, 4 and 5 represent the SEGSNR computed over the training and testing frame as function of the number of epochs, in the ADPCM structure (with Nq= number of quantizer bits =4 bits). Main observations are:

! Initially the SEGSNR grows fastly, and is similar for training and testing frames, being sometimes even better for the test frames, which reveals a very good generalization capability, but if the number of epochs is increased the SEGSNR is reduced, specially for the test frame, because of the network specialization in the training (uncorrupted with quantization noise) frame. Obviously the decrease of SEGSNR is more significative for the test frames. This implies that the number of trained epochs must be more carefully selected for the backward configuration than for the forward configuration.

! Figures 4 and 5 represent a better random weights and biases initialization than the figure 3 because the SEGSNR is greater.

After an empirical study, made with a larger number of experiments, we have found that:

a) The difference between test and train SEGSNR is less significative for a bad initialization.

b) the number of trained epochs is not as critical for a bad initialization as for the good ones.

c) If the number of epochs is very high then the final result for the test frame and the bad initialization is better than the obtained with the good initializations.

d) The best result is obtained with a good random initialization and with a small number of epochs. This fact is interesting because it is also a way of limiting the computational complexity (the lesser the number of epochs the lesser the number of required flops for training the network).

For achieving a good initialization a multi-start algorithm is used, which consists in computing several random initializations (experimentally fixed to 4) and to choose the one that achieves the higher SEGSNR.

For selecting the number of epochs the optimal condition would be to evaluate for each frame the number of epochs that maximizes the SEGSNR.( Figure 5 shows a histogram of the optimal number of epochs for the frames of one sentence.) This is impractical because the decoder needs to know the number of epochs in order to track the encoder. Obviously this would imply the transmission of the number of trained epochs and so, the bit rate would be increased. The adopted solution consisted on a statistical study for choosing the best average number of epochs. This study reveals that the optimal number of epochs is 6.

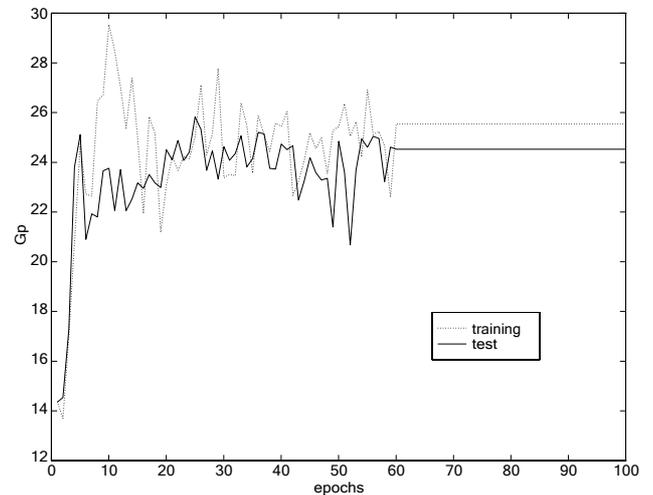

Fig. 3 SEGSNR vs trained epochs for a random initialization

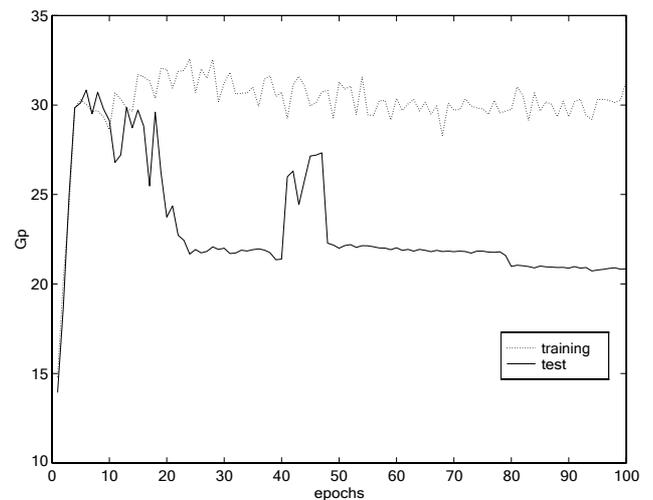

Fig. 4 SEGSNR vs trained epochs for a random initialization

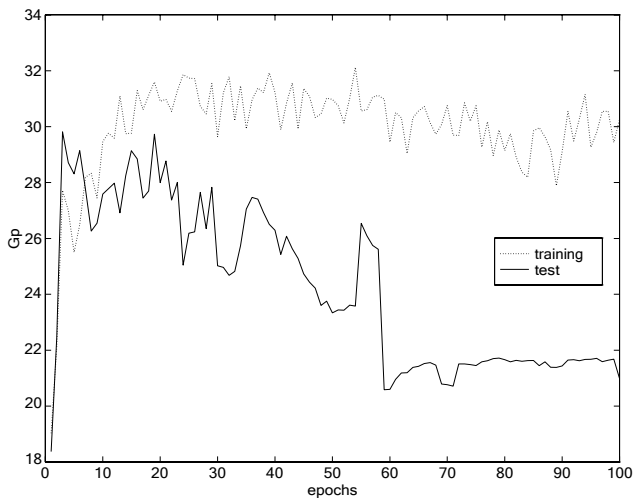

Fig. 5 SEGSNR vs trained epochs for a random initialization

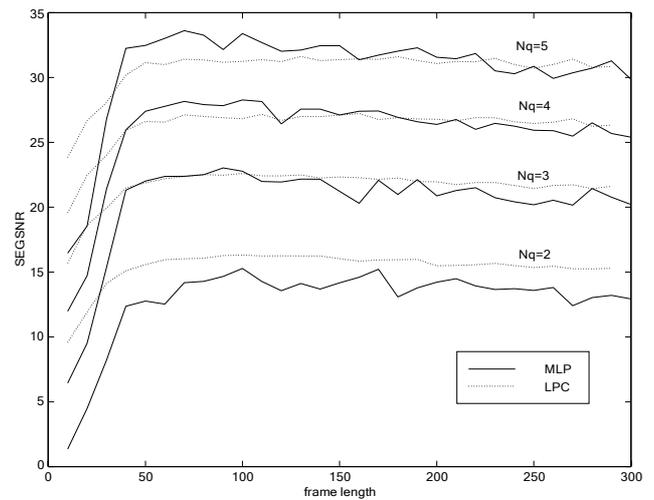

Fig. 7 SEGSNR vs frame length for a female speaker

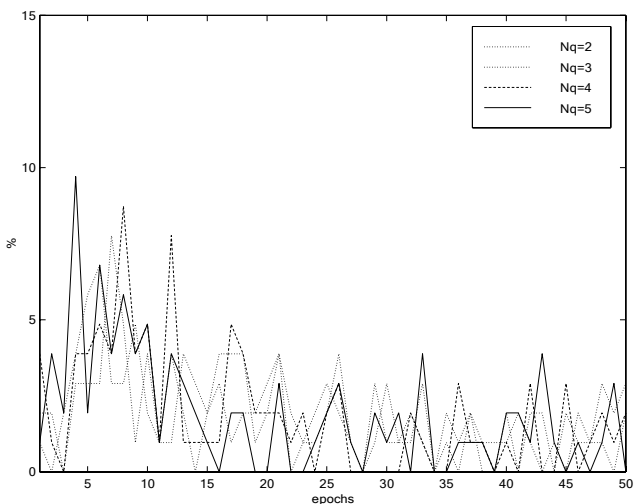

Fig. 6 percentage of frames with obtain the given value of epochs as optimum.

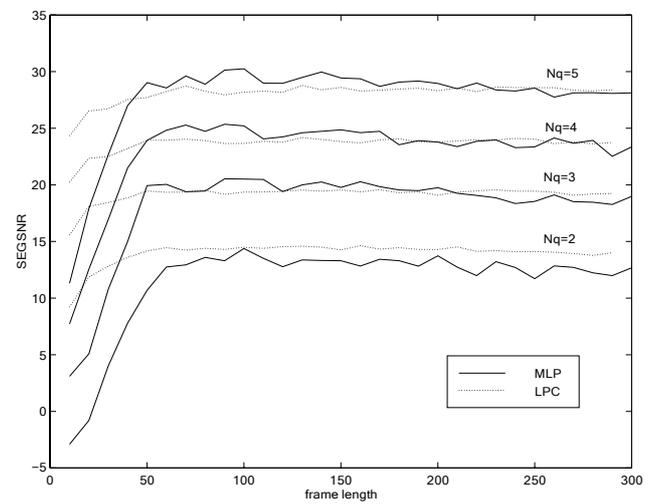

Fig. 8 SEGSNR vs frame length for a male speaker.

! Frame length: Same commentaries of the linear predictor apply here. Experimental results show that the linear predictor has a similar behaviour over a wider range of frame sizes than the nonlinear predictor, but there is some rage for which the nonlinear predictor is better than the linear predictor.

Figures 7 and 8 shows the SEGSNR (computed with a 200 samples analysis window) for frame lengths ranging from 10 to 300 samples for MLP10x2x1 and LPC-10 with Nq=2 to 5 bits, averaged for the frames of one sentence.

Although it is possible to optimize the frame length, it is important to remember that:

   ! The number of flops is increased if the frame size is reduced

   ! For the backward configuration, the transmission rate does not depend on the frame length, but for the forward configuration the predictor parameters must be transmitted and if the frame length is reduced the compression ratio is also reduced.

   ! For the hybrid predictor proposed in section 4, an overhead of 1 bit/frame must be sent, so if the frame length is reduced the compression ratio is also reduced.

For these reasons in this study the block size has been selected to 200 samples/frame, because it is a high used value in other applications, and offers a good compromise.

## 3. RESULTS

The results have been evaluated using subjective criteria (listening to the original and decoded files), and SEGSNR.

The following table shows the SEGSNR obtained with the ADPCM configuration for the whole database with the following predictors: LPC-10, LPC-25 and MLP 10x2x1.

The results of the ADPCM forward (with unquantized predictor coefficients) are also provided such us reference of the backward configuration.

| METHOD | Nq=2 bits | | Nq=3 bits | | Nq=4 bits | | Nq=5 bits | |
|---|---|---|---|---|---|---|---|---|
| | SEGSNR | std | SEGSNR | std | SEGSNR | std | SEGSNR | std |
| ADPCMF-LPC-10 | 14.98 | 4.8 | 20.68 | 5.8 | 25.2 | 6.4 | 29.89 | 6.8 |
| ADPCMF-LPC-25 | 15.25 | 4.8 | 21.11 | 5.7 | 25.84 | 6.4 | 30.49 | 6.9 |
| ADPCMF-MLP | 15.37 | 6.3 | 23.18 | 6.2 | 28.15 | 6.9 | 32.72 | 7.4 |
| ADPCMB-LPC-10 | 14.29 | 5.1 | 19.98 | 6 | 24.83 | 6.4 | 29.45 | 6.7 |
| ADPCMB-LPC-25 | 14.54 | 5.1 | 20.29 | 5.8 | 24.99 | 6.2 | 29.63 | 6.5 |
| ADPCMB-MLP | 13.76 | 5.7 | 20.04 | 6.6 | 25.26 | 6.9 | 30.1 | 7.3 |
| ADPCMB-HYBRID | 15.24 | 5.1 | 21.28 | 6.2 | 26.35 | 6.6 | 31.42 | 6.9 |

This results reveal the superiority of the nonlinear predictor in the forward configuration (2dB aprox. over LPC-25 except for the 2 bit quantizer). This superiority is greater if the quantizer has a high number of levels.

In the backward configuration there is a small SEGSNR decrease with the linear predictor versus the forward configuration. For the nonlinear predictor it is more significative (nearly 3dB), but the SEGSNR is better than LPC-10 except for Nq=2 bits. Also, the variance of the SEGSNR is greater than for the linear predictor, because in the stationary portions of speech the neural net works satisfactorilly well, and for the unvoiced parts the nnet generalizes poorly. For this reason, a hybrid predictor is proposed. Next section describes the scheme.

A significance test for a difference between means (SEGSNR) was done. We found that there was no difference in the SEGSNR of the LPC with 10 or 25 coefficients with a significance of 1% (i.e.

$$z = \frac{|\ SEGSNR_1 - SEGSNR_2\ |}{\sqrt{\dfrac{std_1^2}{n} + \dfrac{std_2^2}{n}}} < 2.5 \text{ a significant difference}$$

with a confidence of 1% was found between the SEGSNR with the MLP and with the LPC.

## 4. ADPCM BACKWARD- HYBRID WAVEFORM CODER

We propose a linear/non linear switched predictor in order to choose always the best predictor and to increase the SEGSNR of the decoded signal. Figure 7 represents the implemented scheme.

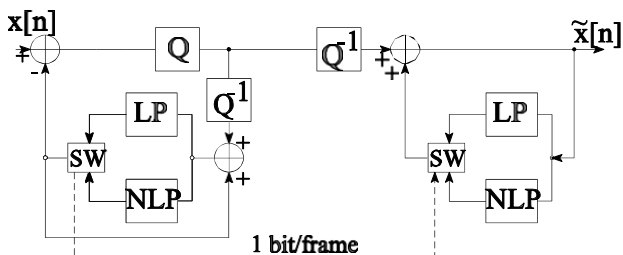

Fig. 9 ADPCM-B hybrid coder. LP: linear predictor, NLP: nonlinear predictor, SW: switch

For each frame the outputs of the linear and nonlinear predictor are computed simultaneously with the coefficients deducted from the previous encoded frame. Then a logical decision is made that chooses the output with smaller prediction error. This implies an overhead of 1 bit for each frame that represents only 1/200 bits more per sample (in our simulations frame size is 200 samples). It is referred in the table as hybrid predictor, because it combines linear and nonlinear technologies. The percentage of use of each predictor is: 55.8% MLP and 44.2% LPC-10.

## 5. CONCLUSIONS AND COMPARISON WITH PREVIOUSLY PUBLISHED WORK

The unique work that we have found that deals with ADPCM with nonlinear prediction is the one proposed by Mumolo et alt. [3]. It has problems of unstability, which were overcome with a switched linear/nonlinear predictor.

Our novel nonlinear scheme has been always stable in our experiments.

The results of our novel scheme show an increase of 1 to 2 dB over classical LPC-10 for quantizer ranges from 2 to 5 bits, while the work of Mumolo [3] is 1 dB over classical LPC for quantizer ranges from 3 to 4 bits and also with and hybrid predictor.

The improvement can be increased if the frame length is decreased to an appropriate value, at the cost of more computational complexity.

### ACKNOWLEDGEMENTS

This work has been supported by the CICYT TIC97-1001-C02-02